\documentclass[10pt,twocolumn,a4paper]{IEEEtran} 
\usepackage{times}
\usepackage{fullpage}
\usepackage{graphicx}

\newcommand{\figuracentrata}[3]
{
\begin{figure*}
  \centering
 \includegraphics[width=12cm]{#1}
  \caption{#2}\label{#3}
\end{figure*}
}

\newcommand{\figura}[3]
{
\begin{figure}
  \centering
 \includegraphics[width=7.5cm]{#1}
  \caption{#2}\label{#3}
\end{figure}
}

\begin{document}
\title{Design of Multistage Decimation Filters Using
Cyclotomic Polynomials:\\Optimization and Design Issues
}
\author{Massimiliano Laddomada,~\IEEEmembership{Member,~IEEE}
\thanks{The author is with the Dipartimento di Elettronica, Politecnico di Torino, Corso Duca
degli Abruzzi 24, 10129 Torino, Italy.
E-mail:{\tt~laddomada@polito.it}}}
\maketitle
\begin{abstract}
This paper focuses on the design of multiplier-less decimation
filters suitable for oversampled digital signals. The aim is
twofold. On one hand, it proposes an optimization framework for
the design of constituent decimation filters in a general
multistage decimation architecture. The basic building blocks
embedded in the proposed filters belong, for a simple reason, to
the class of cyclotomic polynomials (CPs): the first 104 CPs have
a z-transfer function whose coefficients are simply $\{-1,0,+1\}$.
On the other hand, the paper provides a bunch of useful
techniques, most of which stemming from some key properties of
CPs, for designing the proposed filters in a variety of
architectures. Both recursive and non-recursive architectures are
discussed by focusing on a specific decimation filter obtained as
a result of the optimization algorithm.

Design guidelines are provided with the aim to simplify the design
of the constituent decimation filters in the multistage chain.
\end{abstract}
%
%
\section{Introduction and Problem Formulation}
\label{Introduction_Problem_Formulation}
\figuracentrata{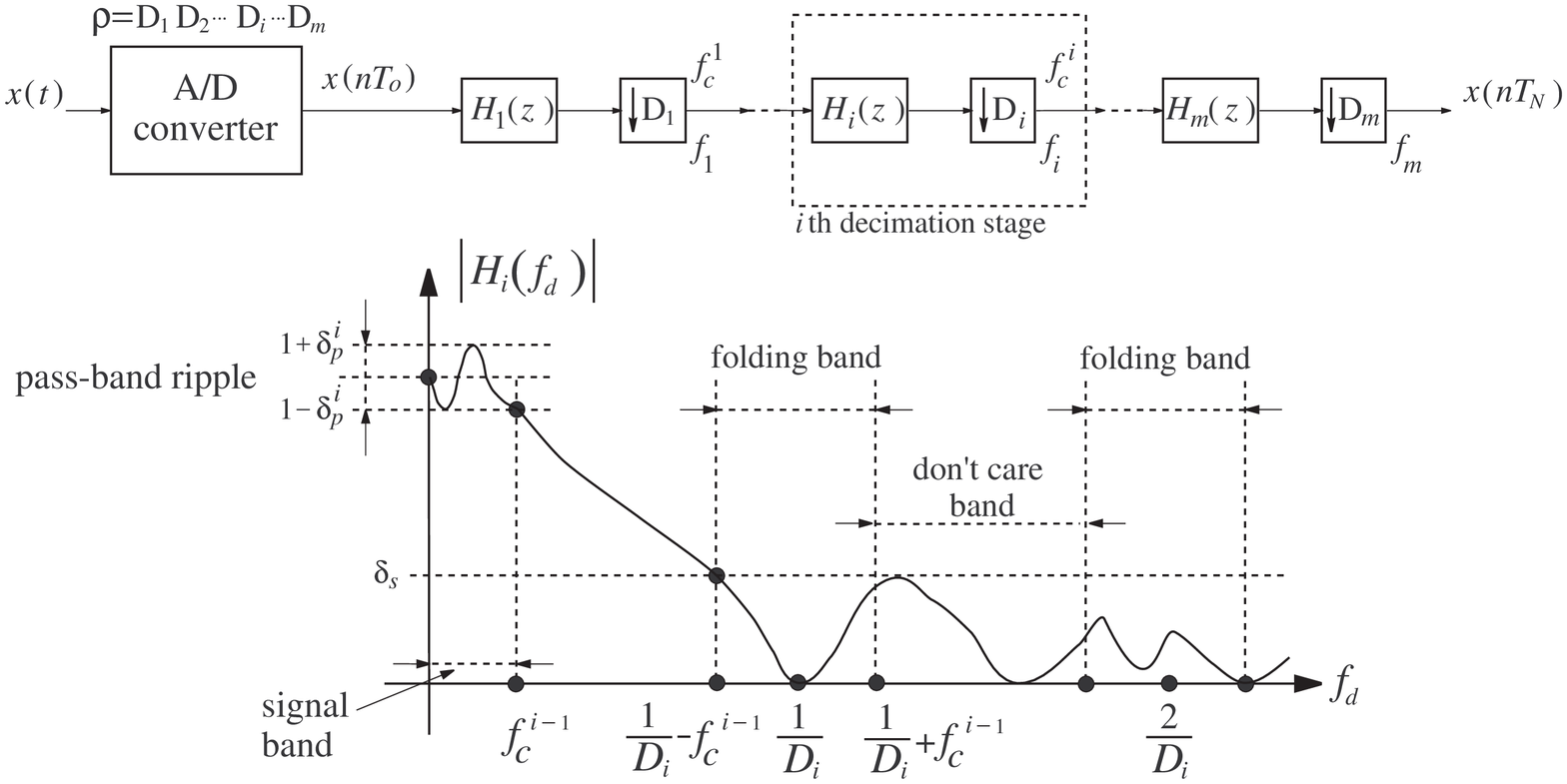}{General
architecture of a $m$-stage decimation chain for A/D converters,
along with a pictorial representation of the key frequency
intervals to be carefully considered for the design of the $i$th
decimation stage. The sampling rate at the input of the $i$th
decimation stage is $f_{i-1},~\forall i=1,\ldots,m$.}{arch}
\noindent The design of multistage decimation filters for
oversampled signals is a well-known research
topic~\cite{CrochiereRabiner}. Mainly inspired by the need of
computationally efficient architectures for wide-band,
multi-standard, reconfigurable receiver design, this research
topic has recently garnered new emphasis in the scientific
community~\cite{Mitola}-\cite{AbidiSR}. Multistage decimation
filters are also employed for decimating highly oversampled
signals from noise-shaping $\Sigma\Delta$ A/D
converters~\cite{Temes}.

Given a base-band analog input signal $x(t)$ with bandwidth
$\left[-B_x,+B_x\right]$, an A/D converter produces a digital
signal $x(nT_o)$ by sampling $x(t)$ at rate
$f_o=\frac{1}{T_o}=2\rho B_x\gg 2B_x$, whereby $\rho\ge 1$ is the
oversampling ratio (notice that $\rho> 1$ for oversampled
signals). The normalized maximum frequency contained in the input
signal is defined as $f^o_c=\frac{B_x}{f_o}=\frac{1}{2\rho}$, and
the digital signal $x(nT_o)$ at the input of the first decimation
filter has frequency components belonging to the range
$[-f^o_c,f^o_c]$. This setup is pictorially depicted in the
reference architecture shown in Fig.~\ref{arch}.

Owing to the condition $\rho\gg 1$, the decimation of an
oversampled signal $x(nT_o)$ is
efficiently~\cite{CrochiereRabiner} accomplished by cascading two
(or more) decimation stages as highlighted in Fig.~\ref{arch}, in
which a multistage architecture composed by $m$ decimation stages
is shown as reference scheme. Consider an oversampling ratio
$\rho$ which can be factorized as follows:
\[
\rho=\prod_{i=1}^{m}D_i
\]
whereby, for any $i$, $D_i$ is an appropriate integer strictly
greater than zero.
%

In the general architecture shown in Fig.~\ref{arch}, sampling
rate decreases in $m$ consecutive stages, whereby the sampling
rate at the input of the $i$th stage is
\[
f_{i-1}=f_{i}\cdot D_{i},~\forall i=1,\ldots,m
\]
while the output sample data rate is:
\[
f_i=\frac{f_o}{\prod_{p=1}^{i}D_p},~\forall i=1,\ldots,m
\]
The design of any decimation stage in a multistage architecture
imposes stringent constraints on the shape of the frequency
response over the so-called folding bands. Considering the scheme
in Fig.~\ref{arch}, the frequency response $H_i(e^{j\omega})$ of
the $i$th decimation filter must attenuate the quantization noise
(QN) falling inside the frequency ranges defined as
\begin{equation}\label{folding_bands_def}
\begin{array}{lll}
\left[\frac{k}{D_i}-f^{i-1}_c;\frac{k}{D_i}+f^{i-1}_c\right],&k=1,...,k_M&\\
k_M=\lfloor \frac{D_i}{2}\rfloor,&D_i~\textrm{even}&\\
k_M=\lfloor \frac{D_i-1}{2}\rfloor,&D_i~\textrm{odd}&
\end{array}
\end{equation}
%
%
%
whereby $f^{i-1}_c$ is the normalized signal bandwidth at the
input of the $i$th decimation filter. The reason is simple: the QN
falling inside these frequency bands will fold down to baseband
(i.e., inside the useful signal bandwidth
$\left[-f^{i-1}_c,+f^{i-1}_c\right]$) because of the sampling rate
reduction by $D_i$ in the $i$th decimation stage, irremediably
affecting the signal resolution after the multistage decimation
chain.

\noindent On the other hand, frequency ranges labelled as
\textit{don't care} bands in Fig.~\ref{arch}, do not require a
stringent selectivity since the QN within these bands will be
rejected by the subsequent filters in the multistage chain.

The relation between $f^i_c$ and $f^o_c$ is as follows:
\[
f^i_c=f^{i-1}_c D_{i},~\forall i=1,\ldots,m
\]
whereby it is $f^o_c=1/2\rho$.

The $i$th decimation filter $H_i(e^{j\omega})$ introduces a
pass-band ripple $\delta^i_p$ which can also be expressed in dB as
follows
\begin{equation}\label{passband_ripple}
R^i_p=-20\log_{10}\left(\frac{1-\delta^i_p}{1+\delta^i_p}\right)>0
\end{equation}
while the selectivity (in dB) corresponds to
\begin{equation}\label{filter_selectivity}
A_s=20\log_{10}\left(\frac{\delta_s}{1+\delta^i_p}\right)\approx
20\log_{10}\left(\delta_s\right)\ll 0
\end{equation}
%
%
%
%
With this background, let us provide a quick survey of the recent
literature related to the problem addressed here. This survey is
by no means exhaustive and is meant to simply provide a sampling
of the literature in this fertile area.

Excellent tutorials on the design of multirate filters can be
found in~\cite{CrochiereRabinerTut,Vaidyanathan}, while an
essential book on this topic is~\cite{CrochiereRabiner}. Recently,
Coffey~\cite{Coffey1,Coffey2} addressed the design of optimized
multistage decimation and interpolation filters.

The design of cascade-integrator comb (CIC) filters was first
addressed in~\cite{Hoge}, while multirate architectures embedding
comb filters have been discussed in~\cite{Chu}. Since then, many
papers \cite{Losada} have focused on the computational
optimization of CIC filters even in the light of new wide-band and
recofigurable receiver design
applications~\cite{gao}-\cite{Ze_Tao}. Comb filters have been then
generalized in~\cite{Letizia}-\cite{laddomada_sharp}, especially
in relation to the decimation of $\Sigma\Delta$ modulated signals.

Other works somewhat related to the topic addressed in this paper
are~\cite{kwentus}-\cite{Yong_Lian}. The use of decimation
sharpened filters embedding comb filters is addressed
in~\cite{laddomada_sharp}-\cite{kwentus}, while
in~\cite{aboushady} authors proposed computational efficient
decimation filter architectures using polyphase decomposition of
comb filters. Dolecek \textit{et al.} proposed a novel two-stage
sharpened comb decimator in~\cite{dolecek}. The design of FIR
filters using cyclotomic polynomial (CP) prefilters has been
addressed in~\cite{Boudreaux_Bartels}, while effective algorithms
for the design of low-complexity FIR filters embedding CP
prefilters have been proposed in~\cite{OhLee2}-\cite{Yong_Lian}.

Owing to the discussion on the folding bands presented above, this
paper addresses the design of computationally efficient decimation
filters suitable for oversampled digital signals. Natural eligible
blocks used in filter design are cyclotomic polynomials with order
less than $105$, since these polynomials possess coefficients
belonging to the set $\{-1,0,+1\}$.
%
%
We first recall the basic properties of CPs in
Section~\ref{CP_basic_properties} since these properties suggest
useful hints at the basis of the practical implementation of the
designed decimation filters. For conciseness, we address the
design of the first stage in the multistage architecture, even
though the considerations which follow are easily applicable to
any other stage in the chain.

The computational complexity of basic CP filters is discussed in
Section~\ref{complexity_of_CPs}. In
Section~\ref{Optimization_algorithm} we propose an optimization
framework whose main aim is to design an optimal decimation filter
(optimal in that the cost function to be minimized accounts for
the number of additions required by the chosen CP filter)
featuring high selectivity within the folding bands seen from the
$i$th decimation stage.

\noindent The practical implementation of the designed decimation
filters is addressed in
Section~\ref{Implementation_Issues_section}, whereby both
recursive and non-recursive architectures stemming from a variety
of properties of polynomials, are discussed. Finally,
Section~\ref{conclusions} draws the conclusions.
%
%
%
%
%
%
\section{Basics of Cyclotomic Polynomials and Key Properties}
\label{CP_basic_properties}
\noindent Cyclotomic polynomials (CPs) arose hand in hand with the
old Greek problem of dividing a circle in equal parts. Key
properties of such polynomials along with the basic rationales can
be found in various number theory books (we invite the interested
readers to refer to~\cite{Schroeder,McClellanRader}), other than
in some recent papers~\cite{Boudreaux_Bartels}. Given an integer
$D$ strictly greater than zero, polynomial $\left(1-z^{-D}\right)$
can be factorized as a product of cyclotomic polynomials as
follows:
\begin{table}
\caption{Values of the totient function for $n\in [1,69]$. Symbol
$\underline{1}$ is used to signify the fact that the underlined
number is associated to $n=1$, while following numbers are
associated to increasing values of $n$.}
\begin{center}
\begin{tabular}{c} \hline

$\phi (n)$
\\\hline \hline

\underline{1}, 1, 2, 2, 4, 2, 6, 4, 6, 4, 10, 4, 12, 6, 8, 8, 16,
6, 18, 8,
12, \\

10, 22, 8, 20, 12, 18, 12, 28, 8, 30, 16, 20, 16, 24, 12, 36, 18,
 \\

24, 16, 40, 12, 42, 20, 24, 22, 46, 16, 42, 20, 32, 24, 52, 18,
\\40, 24, 36, 28, 58, 16, 60, 30, 36, 32, 48, 20, 66, 32, 44

\\\hline\hline

\end{tabular}
\label{totient_values}
\end{center}
\end{table}
\begin{equation}
1-z^{-D}=\prod_{q:q|D}^{D}C_q\left(z\right)
 \label{cycl_pol}
\end{equation}
whereby $q:q|D$ identifies the set of integers $q$, less than, or
equal to $D$, which divides $D$ (in other words, the remainder of
the division between $D$ and $q$ is zero). For each $q$ as above,
there is a unique polynomial $C_q\left(z\right)$ whose roots
satisfy the following conditions.
\begin{itemize}
    \item For each $q\le D$, the roots of $C_q\left(z\right)$ constitute a
    subset of the roots belonging to the polynomial $1-z^{-D}$.
    \item The roots of $C_q\left(z\right)$ are the primitive
    $q$th roots of unity, i.e., they all fall on the $z$-plane
    unit circle.
    \item The number of roots corresponds to the number of
    positive integers which are prime with respect to $D$, and
    smaller than $D$.
    \item Roots of $C_q\left(z\right)$ do not belong to the set of roots of
    the polynomial $1-z^{-r},~\forall r:~ 0<r<q\le D$.
\end{itemize}
Based on the observations above, polynomials $C_q\left(z\right)$
are defined as:
\begin{equation}
C_q\left(z\right)=\prod_{i:(i,q)=1}^{q}\left(1-z^{-1}e^{-j2\pi
\frac{i}{q}}\right)
 \label{c_q}
\end{equation}
whereby $(i,q)=1$ is used to mean that $i$ and $q$ are
co-prime~\cite{Schroeder}. Notice that, given an integer $q$,
(\ref{c_q}) allows us to write the $z$-transfer function of any CP
indexed by $q$.

\noindent Key advantages of CPs in connection to filter design
rely on the following property: if $q$ has no more than two
distinct odd prime factors, polynomials $C_q\left(z\right)$
contains coefficients belonging to the set $\{-1,0,+1\}$. From a
practical point of view, CP coefficients belong to the set
$\{-1,0,+1\}$ if $q\le 104$~\cite{Schroeder,McClellanRader}.

The degree of polynomial $C_q\left(z\right)$ is not $q$ but it is
defined as follows:
\begin{equation}\label{cq_degree}
\textrm{deg}\left[C_q\left(z\right)\right]=\sum_{d|q}d\cdot \mu
\left(\frac{q}{d}\right)=\phi (q)
\end{equation}
whereby $\phi (q)$ is the totient function (see
Table~\ref{totient_values}), i.e., the number of positive integers
less or equal to $q$ that are relatively prime\footnote{Two
numbers are said to be relatively prime if they do not contain any
common factor. Notice that the integer $1$ is considered as being
relatively prime to any integer number.} to $q$, while $\mu
\left(n\right)$ is the M$\ddot{\textrm{o}}$bius function defined
as:
\begin{equation}\label{mu_function}
\mu \left(n\right)=\left\{
      \begin{array}{ll}
       1, &  n=1 \\
       (-1)^k, & n=p_1\cdot p_2 \cdot\ldots\cdot p_k,\\
       &\textrm{with}~ p_i ~\textrm{prime},~p_i\ne
       p_j,~\forall i\ne j\\
       0, & \textrm{if n is divisible}\\
        &\textrm{by the squares of a prime}
      \end{array}
\right.
\end{equation}
Index $k$ in the second entry stands for the number of distinct
prime numbers which decomposes the argument $n$.
Values of the M$\ddot{\textrm{o}}$bius function are shown in
Table~\ref{moebius_values} for $n\in [1,104]$. Notice that $\mu
\left(n\right)\ne 0$ implies that $n$ is squarefree, i.e., its
decomposition does not contain repeated factors.
\begin{table}
\caption{Values of the M$\ddot{\textrm{o}}$bius function for $n\in
[1,104]$.}
\begin{center}
\begin{tabular}{c|l} \hline

$\mu \left(n\right)$ & $n$
\\\hline \hline

$-1$ & 2, 3, 5, 7, 11, 13, 17, 19, 23, 29, 30, 31, 37, 41, 42,  \\

& 43, 47, 53, 59, 61, 66, 67, 70, 71, 73, 78, 79, 83, 89, \\
& 97, 101, 102, 103
\\\hline

$1$ & 1, 6, 10, 14, 15, 21, 22, 26, 33, 34, 35, 38, 39, 46,
 \\

&51, 55, 57, 58, 62, 65, 69, 74, 77, 82, 85, 86, 87, \\
 &91, 93, 94, 95
\\\hline

$0$ & 4, 8, 9, 12, 16, 18, 20, 24, 25, 27, 28, 32, 36, 40, 44, \\

& 45, 48, 49, 50, 52, 54, 56, 60, 63, 64, 68, 72, 75, 76, \\
&80, 81, 84, 88, 90, 92, 96, 98, 99, 100, 104 \\\hline\hline

\end{tabular}
\label{moebius_values}
\end{center}
\end{table}

The $z$-transfer function of a CP with squarefree index $q$ is
\cite{matematica}:
\begin{equation}\label{squarefree_cycl_pol}
C_q(z)=\sum_{d=0}^{\phi (q)}c_{q,d}z^{-(\phi (q)-d)}
\end{equation}
whereby coefficients $c_{q,d}$ can be evaluated with the following
recursive relation:
\begin{equation}\label{squarefree_cycl_pol_2}
c_{q,d}=-\frac{\mu (q)}{d}\sum_{p=0}^{d-1}c_{q,p}\cdot
\mu\left(g(q,d-p)\right)\phi\left(g(q,d-p)\right)
\end{equation}
using the initial value $c_{q,0}=1$. Function $g(q,d-p)$ in
(\ref{squarefree_cycl_pol_2}) is the greatest common divisor
between $q$ and $d-p$. Notice that~(\ref{squarefree_cycl_pol_2})
represents an effective algorithm for automatically generating the
$z$-transfer function of CPs with squarefree indexes $q$.

Perhaps, the main properties useful for deducing the $z$-transfer
function of any CP, are the ones summarized in the
following~\cite{Schroeder}. We will discuss the application of
such properties in Section~\ref{complexity_of_CPs} whereby the
focus is on the design of low complexity CPs in terms of both
additions and delays.
\begin{enumerate}
    \item Given a prime number $t$, it is
\begin{equation}\label{cyclo_prime}
C_t(z)=\sum_{i=0}^{t-1}z^{-i}=\frac{1-z^{-t}}{1-z^{-1}}
\end{equation}
    \item Let $k$, $n$ and $m$ be three positive integers. Then,
    it is
\begin{equation}\label{cyclo_nm}
    C_{mn^k}\left(z\right)=C_{mn}\left(z^{n^{k-1}}\right)
\end{equation}
    \item Consider a prime number $p$, which does not divide $q$,
    then
\begin{equation}\label{cyclo_pq}
    C_{pq}(z)=\frac{C_{q}(z^{p})}{C_{q}(z)}
\end{equation}
    \item Given any odd integer $n$ greater or equal to $3$, then
    it is
\begin{equation}\label{cyclo_n_odd}
    C_{2n}(z)=C_{n}(-z)
\end{equation}
    \item For $z=1$, the following relation holds:
\begin{equation}\label{cyclo_DC_component}
    C_q(1)=\left\{
        \begin{array}{ll}
        0, &  q=1 \\
        p, & q=p^k,~p~\textrm{prime}\\
        1, & \textrm{otherwise}
        \end{array}
    \right.
\end{equation}
This relation assures us that for indexes $q>1$, $z$-transfer
function of the respective CP presents unity gain in baseband
provided that $q\ne p^k$. Otherwise, CP transfer functions have to
be normalized by $p$ in order to assure unity gain in baseband.
\end{enumerate}
\section{Criteria for Identifying Low Complexity CPs}
\label{complexity_of_CPs}
\noindent The $z$-transfer function of CPs for any index $q$ can
be deduced upon employing the relation~(\ref{c_q}) along with the
properties stated in~(\ref{cyclo_prime})-(\ref{cyclo_n_odd}).
Different architectures (both recursive and non recursive) for
implementing each CP can be obtained, mainly differing in the
number of additions and delays required. For conciseness, in this
paper we show the $z$-transfer functions of the first sixty CPs in
Table~\ref{some_cyclot_polyn}; the $z$-transfer functions of
$C_q(z)$ for any $q\in\{1,\ldots,104\}$ in both non recursive and
recursive (if any) form can be found in \cite{laddomada_CP_docum}.

Let us discuss some key examples by starting from CP $C_{33}(z)$.
Considering that $33$ is squarefree and given that $p=33$ can be
written as $3\times 11$, whereby $3$ and $11$ are coprimes, there
are three possible architectures for implementing such a
polynomial. The first one stems from~(\ref{squarefree_cycl_pol})
and~(\ref{squarefree_cycl_pol_2}) and it consists of a non
recursive architecture (see Table~\ref{some_cyclot_polyn})
employing $14$ additions and $20$ delays. On the other hand, two
recursive architectures follow upon using property
(\ref{cyclo_pq}) with $p=3,q=11$ and $p=11,q=3$:
\begin{equation}\label{C_33_architectures}
\begin{array}{llll}
C_{11\cdot 3}(z)&=\frac{C_{11}(z^{3})}{C_{11}(z)}&= \frac{1-z^{-33}}{1-z^{-3}}\cdot \frac{1-z^{-1}}{1-z^{-11}}&\\
&&=\frac{1-z^{-1}-z^{-33}+z^{-34}}{1-z^{-3}-z^{-11}+z^{-14}}&\\
C_{3\cdot
11}(z)&=\frac{C_{3}(z^{11})}{C_{3}(z)}&=\frac{1+z^{-11}+z^{-22}}{1+z^{-1}+z^{-2}}&
\end{array}
\end{equation}
%
%
As far as the number of additions is concerned, from
(\ref{C_33_architectures}) it easily follows that the architecture
$C_{3\cdot 11}(z)$ only requires 4 additions, which compares
favorably with both the non recursive implementation and
$C_{11\cdot 3}(z)$. Notice also that, since CP coefficients are
simply $\{-1,0,+1\}$, the recursive architectures can be
implemented without coefficient quantization; this in turn
suggests that exact pole-zero cancellation is not a concern with
these architectures.

On the other hand, the non recursive architecture requires only
$20$ delays as opposed to the recursive architectures requiring,
respectively, $34$ and $22$ delays. In this work, we suppose that
the computational complexity of the filter depends only on the
number of additions.

Upon comparing for any $q$ both recursive and non recursive
architectures in Table~\ref{some_cyclot_polyn} (see also the
complete list of the first CPs reported
in~\cite{laddomada_CP_docum}), it easily follows that recursive
implementations, when do exist, allow the reduction of the number
of additions with respect to non recursive implementations; the
price to pay, however, relies on the increased filter delay. As a
rule of thumb, non recursive architectures should be preferred to
recursive implementations when memory space is a design
constraint. On the other hand, recursive architectures can greatly
reduce the number of additions.

Let us briefly discuss the possible architectures related to an
even indexed CP, such as $C_{60}(z)$. By virtue of the different
ways to factorize the integer $60$, property~(\ref{cyclo_pq}) can
be applied with the following combinations $p=5,q=12$, $p=3,q=20$
whereby in both cases $p$ is a prime integer not dividing $q$.
Property~(\ref{cyclo_nm}) can be applied with $m=15,n=2,k=2$. In
Table~\ref{some_cyclot_polyn} we show only both the recursive and
the non recursive architectures yielding the lowest complexities.

When $q$ is a prime number, the $z$-transfer function of the
related CP corresponds to the first order comb filter, as can be
straightforwardly seen from~(\ref{cyclo_prime}). Finally,
property~(\ref{cyclo_n_odd}) can be effectively employed for
deducing the $z$-transfer function of CPs with even indexes $q$
which can be written as $2n$, with $n$ an odd number strictly
greater than $2$. As an example, notice the following relations:
$C_{30}(z)=C_{15}(-z)$, $C_{34}(z)=C_{17}(-z)$.

The simple examples presented above are by no means a complete
picture of the capabilities and sophistication that can be found
in multistage structures for sampling rate conversion. They are
merely intended to show why such structures can constitute the
starting point for obtaining computationally efficient filters for
decimating oversampled signals. The design of computationally
efficient decimation filters relies on the combination of an
appropriate set of CPs. In oversampled A/D converters, for
example, it is very important to contain the computational burden
of the first stages in the multistage decimation chain. This
motivates the study of an effective algorithm for identifying an
appropriate set of CPs that, cascaded, is able to attain a set of
prescribed requirements as specified in~(\ref{passband_ripple})
and~(\ref{filter_selectivity}): this is the topic addressed in the
next section.
\section{Optimization Algorithm and Design Examples}
\label{Optimization_algorithm}
\begin{table*}\caption{Optimization results}
\begin{center}
\begin{tabular}{c|l}\hline
\hline $D=8$ & Set of eligible CPs: $2,     4,     8,     9, 11,
15, 17,    18,    19,    21,    22,    25,    27,    29,    30,
31, 33, 34, 35, 36,    37,    38,    39,    41,    $\\
&$ 42, 43,44, 45, 47, 49, 50, 51,    53,    54, 55,    57,    58,
59, 60, 61, 62, 63, 64$\\\hline

$A_s=40,~R_p=1$dB & $H_{D8,1}(z)=C_2(z)C_4(z)C^2_8(z)C_{11}(z)$\\

$A_s=50,~R_p=1$dB & $H_{D8,2}(z)=C^2_2(z)C^3_4(z)C^3_8(z)$\\

$A_s=60,~R_p=1$dB &
$H_{D8,3}(z)=C^2_2(z)C^3_4(z)C^3_8(z)C_9(z)$\\\hline

$A_s=40,~R_p=2$dB & $H_{D8,4}(z)=C_4(z)C_8(z)C_{11}(z)C_{17}(z)$\\

$A_s=50,~R_p=2$dB & $H_{D8,5}(z)=C_2(z)C^2_4(z)C^2_8(z)C_{19}(z)$\\

$A_s=60,~R_p=2$dB &
$H_{D8,6}(z)=C_2(z)C_4(z)C^2_8(z)C_{11}(z)C_{17}(z)$\\\hline\hline

%
%
%
%
%
%
%
%

$D=16$ & Set of eligible CPs: $2,     4,     8,     9,    11, 15,
16, 17,    18,    19,    21,    22,    25,    27,    29,    30,
31, 33, 34, 35,    36,    37,    38,    39,    $\\

& $41,    42, 43, 44, 45, 47, 49, 50,    51,    53, 54,    55,
57,    58, 59, 60, 61, 62, 63, 65, 66,    67,    68,    69,    70,
71, 72, 73,
$\\

&$ 74, 75, 76, 77, 78, 79, 81,    82,    83,    84,    85, 86, 87,
88, 89, 90, 91, 93, 94, 95, 97,    98,    99,   100, 101, 102,
103$\\\hline

$A_s=40,~R_p=1$dB & $H_{D16,1}(z)=C_8(z)C_{16}(z)C_{17}(z)C_{19}(z)$\\

$A_s=50,~R_p=1$dB & $H_{D16,2}(z)=C_{11}(z)C_{16}(z)C^2_{17}(z)$\\

$A_s=60,~R_p=1$dB &
$H_{D16,3}(z)=C_4(z)C_{16}(z)C^3_{17}(z)$\\\hline

$A_s=40,~R_p=2$dB & $H_{D16,4}(z)=C_{16}(z)C^2_{29}(z)$\\

$A_s=50,~R_p=2$dB & $H_{D16,5}(z)=C_8(z)C_{16}(z)C_{17}(z)C_{41}(z)$\\

$A_s=60,~R_p=2$dB &
$H_{D16,6}(z)=C_{16}(z)C^2_{17}(z)C_{37}(z)$\\\hline\hline

$D=32$ & Set of eligible CPs:$2,     4,     8,     9,    11, 15,
16, 17,    18,    19,    21,    22,    25,    27,    29,    30,
31, 32, 33, 34, 35,    36,    37,    38,
$\\

& $  39,    41,    42, 43, 44, 45, 47, 49, 50,    51,    53, 54,
55,    57,    58, 59, 60, 61, 62, 63, 65, 66,    67,    68,    69,
70,    71,  $\\

&$72, 73, 74, 75, 76, 77, 78, 79, 81,    82,    83,    84,    85,
86, 87, 88, 89, 90, 91, 93, 94, 95, 97,    98,    99,$\\
& $  100, 101, 102, 103$\\\hline

%
$A_s=40,~R_p=1$dB & $H_{D32,1}(z)=C^2_{31}(z)C_{41}(z)$\\

%
$A_s=50,~R_p=1$dB & $H_{D32,2}(z)=C_{25}(z)C^3_{31}(z)$\\

%
$A_s=60,~R_p=1$dB & $H_{D32,3}(z)=C^4_{17}(z)$\\\hline

$A_s=40,~R_p=2$dB & $H_{D32,4}(z)=C_{31}(z)C_{53}(z)C_{67}(z)$\\

%
$A_s=50,~R_p=2$dB & $H_{D32,5}(z)=C_{16}(z)C^2_{31}(z)C_{79}(z)$\\

%
$A_s=60,~R_p=2$dB &
$H_{D32,6}(z)=C^2_{17}(z)C_{37}(z)C_{67}(z)$\\\hline\hline

\end{tabular}
 \label{Risultati_opt}
\end{center}
\end{table*}
\noindent This section presents an optimization framework for
designing low complexity decimation filters, $H_i(z)$, as a
cascade of CP subfilters. For the derivations which follow,
consider the design of the $i$th decimation filter in the
multistage chain depicted in Fig.~\ref{arch}, with a frequency
response that can be represented as follows:
\begin{equation}\label{freq_resp_H_i}
H_i\left(f_d\right)=\prod_{q=1}^{|S_{cp}|}C^{m_q}_q(f_d)
\end{equation}
whereby $f_d$ is the digital frequency normalized with respect to
the sampling frequency $f_{i-1}$ as discussed in
Section~\ref{Introduction_Problem_Formulation}, $S_{cp}$ is a
suitable set of eligible CPs to be used in the optimization
framework ($|S_{cp}|$ is the cardinality of the set, i.e., the
number of eligible CPs), $C_q(f_d)$ is the frequency response of
the CP indexed by $q$ and $m_q$ is its integer order in the
cascade constituting $H_i\left(f_d\right)$ (it is $m_q\ge
0,~\forall q$).

A suitable cost function accounting for the complexity of the
$i$th decimation filter can be defined as a weighted combination
of the number of adders and delays required by the overall filter
$H_i(z)$~\cite{OhLee}:
\begin{equation}\label{power_consumption}
F\left(m_1,m_2,\ldots,m_{|S_{cp}|}\right)=\sum_{q=1}^{|S_{cp}|}m_q\cdot\left(
N_{a,q}+\gamma \cdot N_{d,q}\right)
\end{equation}
whereby $N_{a,q}$ and $N_{d,q}$ are, respectively, the number of
adders and delays of CP $C_q(z)$, and $\gamma\in [0,1]$ is a
factor depending on the relative complexity of the delays with
respect to the adders. In our setup, we assume that the
computational complexity of the $i$th decimation filter is mainly
due to the number of adders; therefore, we set $\gamma=0$. Notice
that the cost function depends on the CP orders
$m_1,\ldots,m_{|S_{cp}|}$, while $N_{a,q}$ and $N_{d,q}$ are known
once the set $S_{cp}$ of eligible CPs has been appropriately
identified. Notice also that $N_{a,q}$ and $N_{d,q}$ can be
straightforwardly obtained by Table~\ref{some_cyclot_polyn} (see
also~\cite{laddomada_CP_docum} for a list of all 104 CPs).

Let us address the choice of the eligible CPs in the set $S_{cp}$.
This is one of the most important design step since the complexity
of the optimization framework discussed below, is tied tightly to
the number of eligible CPs. By virtue of the discussion on the
folding bands spanned by the $i$th decimation filter, we choose
the eligible CPs between the $104$ CPs in such a way that 1) at
least 20\% of zeros falls within the folding bands defined
in~(\ref{folding_bands_def}), 2) no zero falls in the signal
pass-band ranging from $0$ to $f^o_{c}$. As a result of extensive
tests, we adopted such a threshold which is capable to reject
about $20-60$ initial CPs depending on $D$. Of course, lower
thresholds can increase the number of eligible CPs at the cost of
an increased complexity of the optimization framework discussed
below. On the other hand, when designing the $i$th decimation
filter in a multistage architecture, only the so-called folding
bands must be spanned by zeros, since \textit{don't care}
frequency bands will be appropriately spanned by the zeros
belonging to the subsequent decimation filters in the cascade.

Before presenting the optimization algorithm, let us discuss the
requirements imposed to the frequency response $H_i(f_d)$ of the
$i$th decimation filter in the cascade. Mask
specifications~\cite{CrochiereRabiner} are given as for classical
filters as far as the passband ripple is concerned. In particular,
for the optimization algorithm we use the passband ripple
expressed in dB as specified in~(\ref{passband_ripple}). The main
difference between the design proposed in this work and classical
FIR filter design techniques relies on the fact that in our setup
specifications are only imposed in the folding
bands~(\ref{folding_bands_def}). To this end, we evaluated the
lowest attenuations (worst-case) attained by each CP belonging to
$S_{cp}$ in each folding band:
\begin{equation}
\begin{array}{ll}
A_{d_q}=-\max_{f_d\in
\left[0;+f^{i-1}_c\right]}20\log_{10}\left(|C_q(f_d)|_n\right)   &\\
&\\
A_{s}(k,q)=\min_{f_d\in
\left[\frac{k}{D_i}-f^{i-1}_c;\frac{k}{D_i}+f^{i-1}_c\right]}20\log_{10}\left(|C_q(f_d)|_n\right)&
\end{array}
\end{equation}
whereby subscript $n$ signifies the fact that each CP $C_q$ has
been normalized in such a way as to have unity gain in baseband.
Notice that normalization factors can be deduced
from~(\ref{cyclo_DC_component}). $A_{s}(k,q)$ is the worst
attenuation of the $q$th CP in $S_{cp}$ within the $k$th folding
band, with $k\in \{1,\ldots,k_M\}$, and $k_M$ defined
in~(\ref{folding_bands_def}). Such values (in dB) have been stored
in look-up tables.

Once the set $S_{cp}$ of eligible CPs along with the appropriate
specifications (passband ripple and folding band attenuations)
have been identified, the optimization problem can be formulated
as follows:
\begin{equation}\label{optim_problem}
\begin{array}{lll}
\min_{m_1,\ldots,m_{|S_{cp}|}} F\left(m_1,\ldots,m_{|S_{cp}|}\right)\left|_{\gamma=0}\right.~\textrm{in}~(\ref{power_consumption})& &\\
\textrm{subject~to:}& &\\
%
%
\begin{array}{llll}
0)&\sum_{q=1}^{|S_{cp}|}m_qA_{d_q}&\le   & R_p~\textrm{(ripple)}\\
1)&\sum_{q=1}^{|S_{cp}|}m_qA_{s}(1,q)&\le   & A_s~\textrm{(selectivity)}\\
&\ldots &\ldots &\\
k)&\sum_{q=1}^{|S_{cp}|}m_qA_{s}(k,q)&\le   & A_s\\
&\ldots &\ldots &\\
k_M)&\sum_{q=1}^{|S_{cp}|}m_qA_{s}(k_M,q)&\le   & A_s\\
\end{array}
\end{array}
\end{equation}
The optimization problem can be also solved for different
prescribed selectivities, $A_s$ (as specified
in~(\ref{filter_selectivity})), around the various folding bands.
In this work we do not pursue this approach. However, notice that
such an approach can be effective for noise shaping $\Sigma\Delta$
A/D converters which present an increasing noise power spectra
density for higher and higher values of the digital frequency
$f_d<1/2$~\cite{Temes,laddomada_gcf}. Setting increasing values of
$|A_s|$ in correspondence of successive folding bands can mitigate
noise folding due to the decimation process.

The solution to the optimization problem~(\ref{optim_problem}) is
the set of CP orders $\textbf{m}=[m_1,\ldots,m_{|S_{cp}|}]^T$,
whereby $m_i=0$ signifies the fact that the $i$th CP in $S_{cp}$
is not employed for synthesizing $H_i(f_d)$.

Upon collecting the set of $k_M+1$ conditions in the matrix
$\textbf{A}$:
\[
\textbf{A}=\left(\begin{array}{ccccc}
  A_{d_1} &  \ldots & A_{d_{|S_{cp}|}} \\
  A_{s}(1,1) &  \ldots  & A_{s}(1,|S_{cp}|) \\
  \ldots & \ldots  & \ldots \\
  A_{s}(k_M,1) &  \ldots  & A_{s}(k_M,|S_{cp}|)\\
\end{array}\right)
\]
and the requirements $\textbf{b}=[R_p~ A_s~\ldots~ A_s]^T$, the
constraints in~(\ref{optim_problem}) can be rewritten as follows:
\[
\textbf{A}\textbf{m}\le \textbf{b}
\]
By this setup, the optimization problem in~(\ref{optim_problem})
with respect to $m_1,\ldots,m_{|C^S_p|}$ can be rewritten as
\[
\begin{array}{lll}
\min_{m_1,\ldots,m_{|S_{cp}|}} F\left(m_1,\ldots,m_{|S_{cp}|}\right)\left|_{\gamma=0}\right.& &\\
\textrm{subject~to:}& &\\
%
%
\begin{array}{lll}
&\textbf{A}\textbf{m}\le \textbf{b}&\\
&m_i\ge 0,~m_i~ \textrm{integer},&\forall i=1,\ldots,|S_{cp}|\\
\end{array}
\end{array}
\]
and solved by mixed integer linear programming
techniques~\cite{Papadimitriou}. We solved the optimization
problem using the Matlab function \textit{linprog} along with a
new matlab file capable of managing integer constrained solutions
(the latter file is available online~\cite{MLIP_docum}).

The results of the previous optimization problem are summarized in
Table~\ref{Risultati_opt} for various $A_s$ specifications and two
different values of $R_p$, namely $R_p=1$ and $2$~dB. We solved
the problem for three different values of the decimation factor
$D$ of the first stage in the decimation chain depicted in
Fig.~\ref{arch} by assuming that the residual decimation factor is
$\nu=4$ (in other words, we assumed that $\rho=D\cdot 4$). Notice
that such an approach is quite usual in practice in that the first
decimation filter accomplishes the highest possible decimation in
order to reduce the sampling rate, while the subsequent decimation
stages are usually accomplished with half-band filters each one
decimating by $2$~\cite{CrochiereRabiner}.
\figura{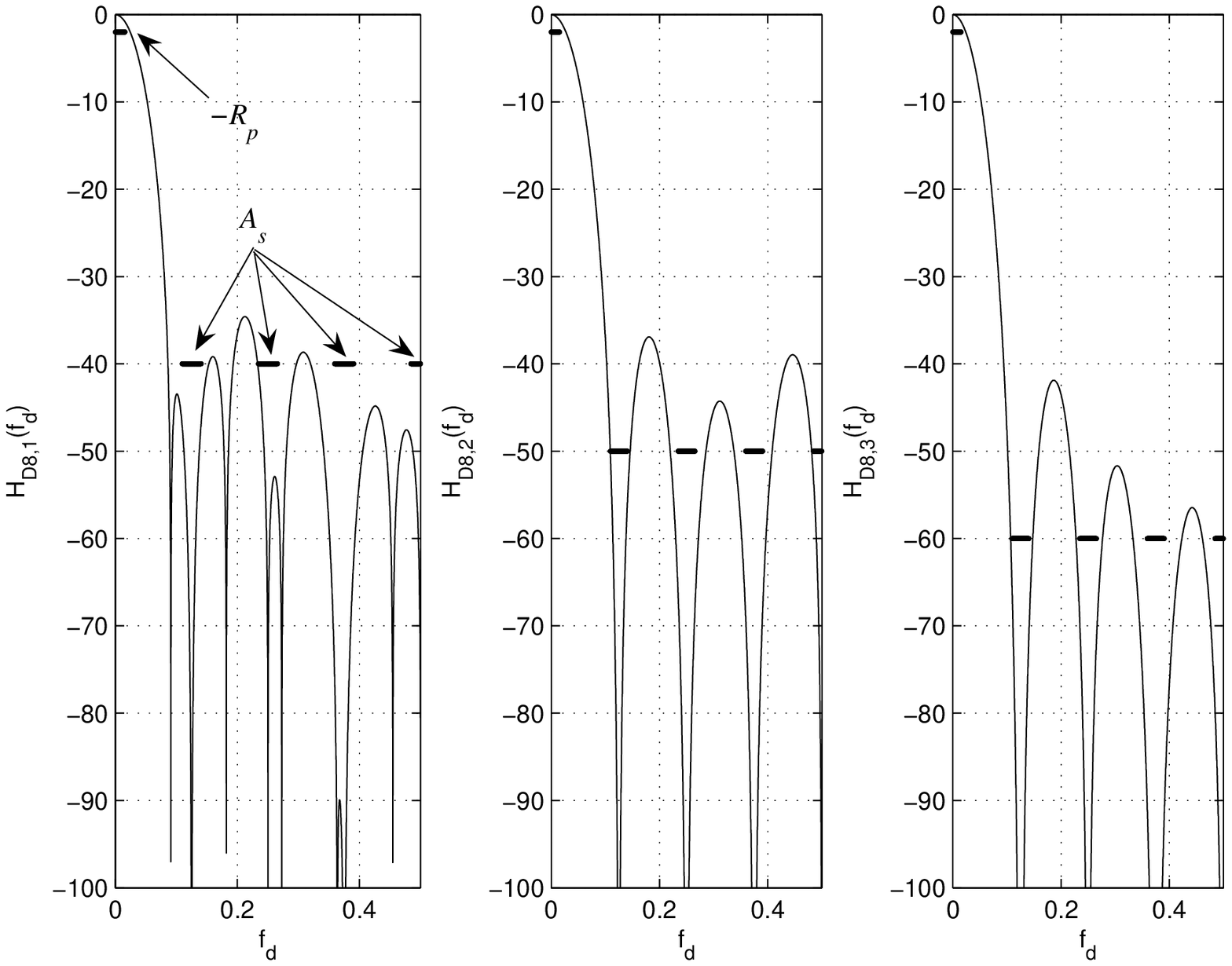}{Behaviours in dB of the modulo of the frequency
responses $H_{8,1}(f_d),H_{8,2}(f_d),H_{8,3}(f_d)$ of the
optimized decimation filters shown in Table~\ref{Risultati_opt}
for $D=8$.}{D8}

The first row related to any decimation factor shows the set of
eligible CPs found in the preliminary design step discussed above,
while the $z$-transfer functions of the CPs can be found in
Table~\ref{some_cyclot_polyn} (see also~\cite{laddomada_CP_docum}
for a list of all 104 CPs).

It is worth comparing the frequency responses of the optimized
filters $H_{8,i}(f_d)$ and $H_{16,i}(f_d)$ (for $i=1,2,3$) in
Table~\ref{Risultati_opt} with the specifications $R_p=1$dB and
various $A_s$. To this end, Fig.s~\ref{D8} and~\ref{D16} show,
respectively, the behaviours of the frequency responses
$H_{8,i}(f_d)$ and $H_{16,i}(f_d)$ along with the imposed
selectivity $A_s$ around the various folding bands (identified by
horizontal bold lines).
%
%
%
%
%
%
%
%
%
%
%
%
\section{Implementation Issues}
\label{Implementation_Issues_section}
\figura{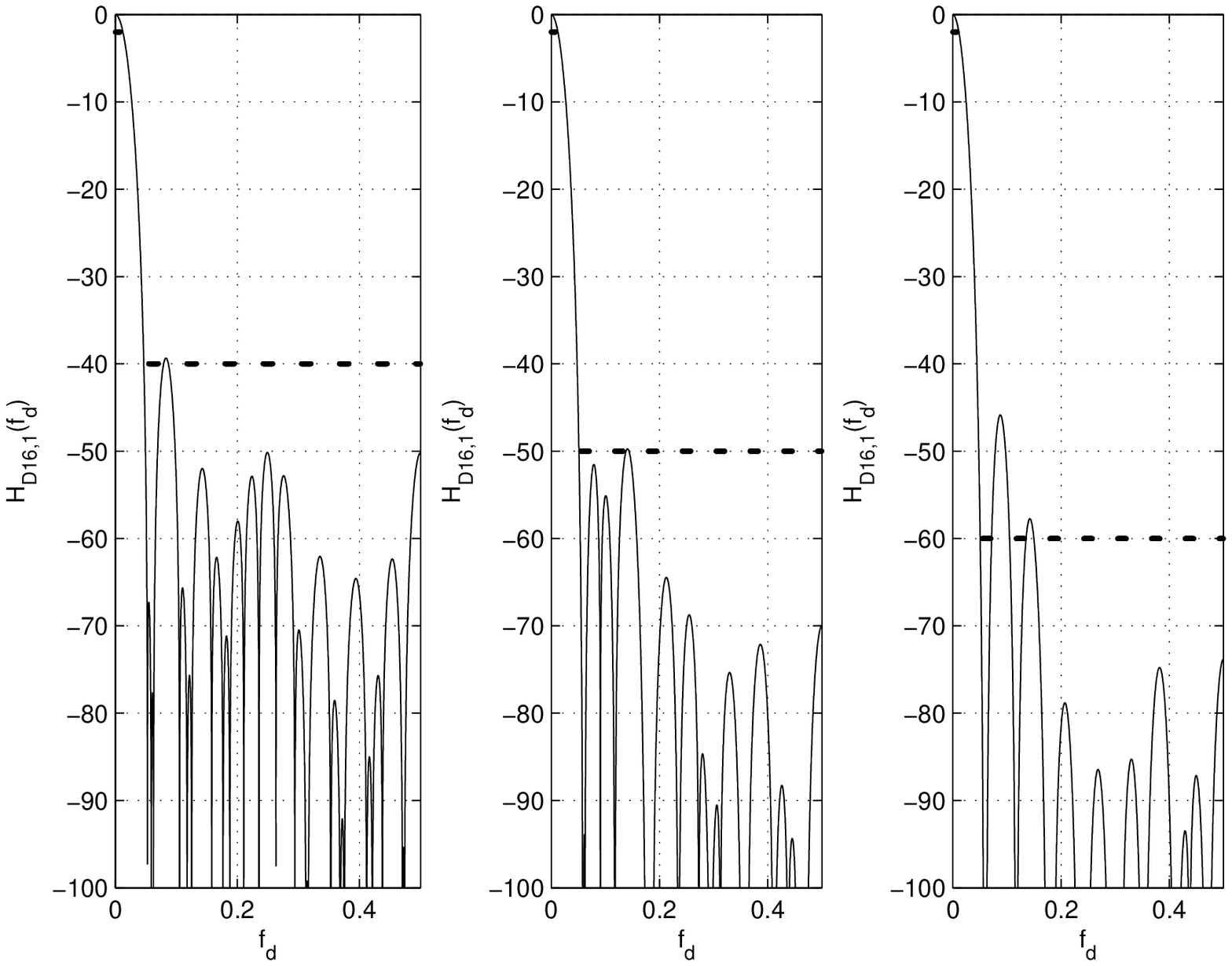}{Behaviours in dB of the modulo of the frequency
responses $H_{16,1}(f_d),H_{16,2}(f_d),H_{16,3}(f_d)$ of the
optimized decimation filters shown in Table~\ref{Risultati_opt}
for $D=16$.}{D16}
\figuracentrata{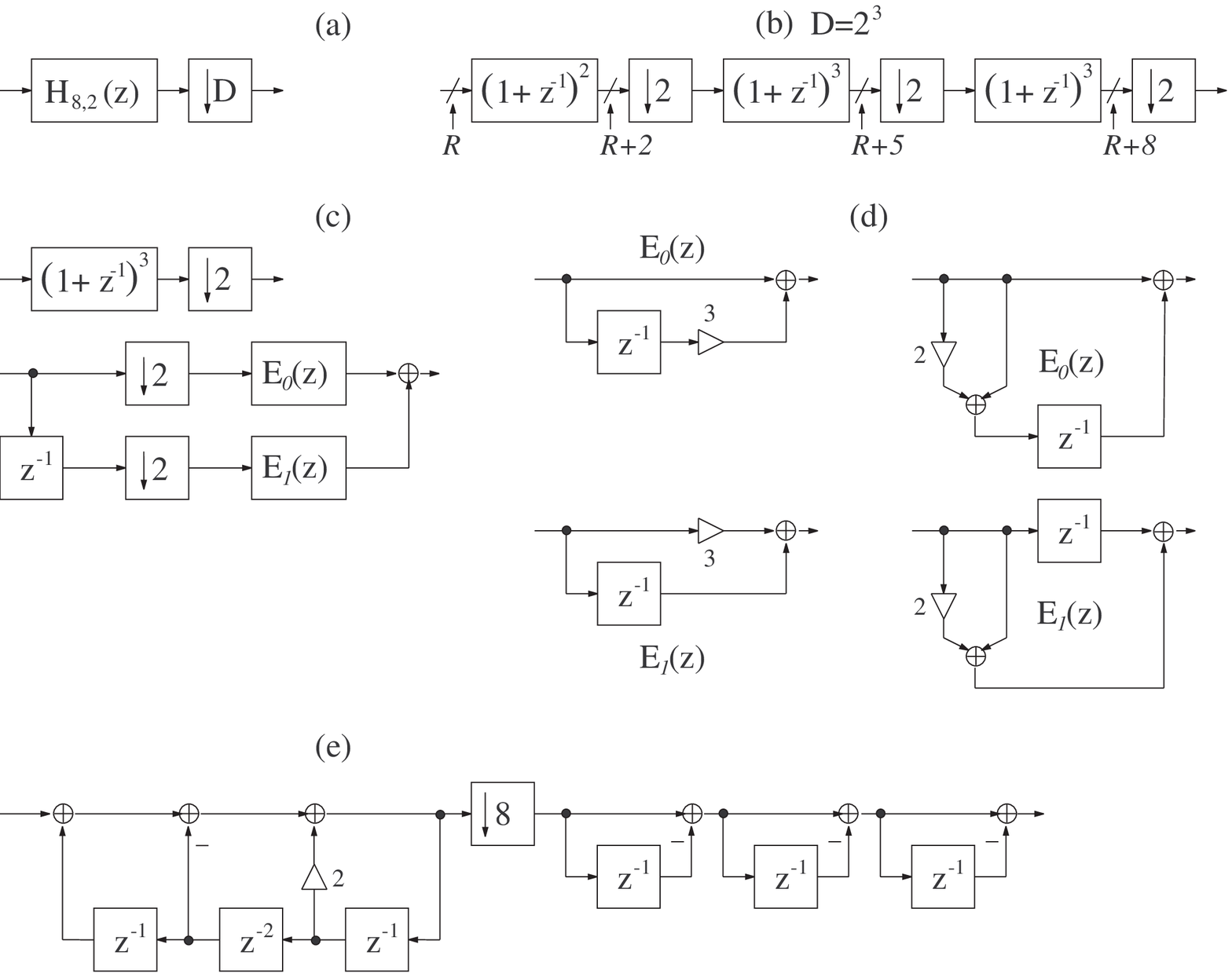}{Efficient architectures for
implementing the decimation stage embedding $H_{8,2}(z)$ (a). Non
recursive architecture (b); polyphase implementation of the
decimation stages decimating by $2$ (c), and polyphase component
implementation using shift registers (d); recursive architecture
of the decimation filter $H_{8,2}(z)$ (e).}{n_rec_impl_gcf3}
\noindent This section addresses the design of optimized CP-based
decimation filters. For conciseness, we will focus on the design
of decimation filter $H_{8,2}(z)$ shown in Table
\ref{Risultati_opt}, even though the considerations which follow
can be applied to any other decimation filter quite
straightforwardly. The decimation stage related to $H_{8,2}(z)$ is
depicted in Fig.~\ref{n_rec_impl_gcf3}a: this decimation filter
will be designed through a variety of architectures following from
different mathematical ways to simplifies the analytical relation
defining $H_{8,2}(z)$.

First of all, notice that upon substituting the appropriate
equations of the constituent CP filters in $H_{8,2}(z)$, the
designed filter takes on the following expression:
\begin{eqnarray}\label{H16_2_z}
H_{8,2}(z)&=& C_2^2(z)C_4^3(z)C_8^3(z)=\\
&=&\left(1+z^{-1}\right)^2\left(1+z^{-2}\right)^3\left(1+z^{-4}\right)^3\nonumber
\end{eqnarray}
which can be rewritten as follows:
\begin{eqnarray}\label{H16_2_z_2}
H_{8,2}(z)&=&
\frac{\prod_{i=0}^{2}\left(1+z^{-2^i}\right)^3}{1+z^{-1}}
\end{eqnarray}
%
%
%
From the commutative property employed in \cite{Chu}, the cascaded
implementation shown in Fig.~\ref{n_rec_impl_gcf3}b easily
follows. The $r$th stage in Fig.~\ref{n_rec_impl_gcf3}b operates
at the sampling rate $f_{i-1}/2^{r}$, whereby $f_{i-1}$ is the
data sampling frequency at the filter input as shown in the
multistage architecture in Fig.~\ref{arch}. Further power
consumption reduction can be achieved by applying polyphase
decomposition to the architecture shown in
Fig.~\ref{n_rec_impl_gcf3}b. To this aim, consider the
$z$-transfer function of the $3$rd order cell:
\begin{eqnarray}\label{Poly3rdStage}
\left(1+z^{-1}\right)^3&=&1+3z^{-2}+z^{-1}\left(3+z^{-2}\right)\nonumber\\
&=&E_0(z^2)+z^{-1}E_1(z^2)\nonumber\\
E_0(z)&=&1+3z^{-1}\nonumber\\
E_1(z)&=&3+z^{-1}
\end{eqnarray}
The polyphase architecture for $\left(1+z^{-1}\right)^3$ easily
follows from the commutative property applied to the two filters
$E_0(z^2)$ and $E_1(z^2)$ in~(\ref{Poly3rdStage}), and it is shown
in Fig.~\ref{n_rec_impl_gcf3}c along with the architectures for
implementing both $E_0(z)$ and $E_1(z)$. Notice that the
multipliers appearing in $E_0(z)$ and $E_1(z)$ can be implemented
in the form of shift registers as depicted in
Fig.~\ref{n_rec_impl_gcf3}d.

The actual complexity of the architecture shown in
Fig.~\ref{n_rec_impl_gcf3}b is fully defined once the data
wordlength in any substage is well characterized, since the power
consumption of a filter cell can be approximated as the product
between the data rate, the number of additions performed at that
rate, and the data wordlength. While the data rate along with the
number of additions are well defined, data wordlength in each
substage in Fig.~\ref{n_rec_impl_gcf3}b is not. Given the input
data wordlength, $R$ (in bits), the data size at the output of the
first decimation substage in Fig.~\ref{n_rec_impl_gcf3}b is equal
to $R+2$ bits since two carry bits have to be allocated for the
two additions involved in that substage. With a similar reasoning,
data wordlength increases at the output of each subsequent
substage in Fig.~\ref{n_rec_impl_gcf3}b in order to take into
account the increase of data size due to the involved additions.

As a reference example, if the decimation filter depicted in
Fig.~\ref{n_rec_impl_gcf3}b is the first decimation stage at the
output of a $\Sigma\Delta$ A/D converter embedding a $1$-bit
quantizer into the loop, it is $R=1$. Thus, data wordlength is as
low as $3$ bits after the first decimation substage, and so on.

Let us address the design of a recursive architecture for
$H_{8,2}(z)$ in~(\ref{H16_2_z}). First of all, consider the
following equality chain
\begin{equation}\label{formula_comb_1_ordine_2}
\prod_{i=0}^{\log_2(D)-1}\left(1+z^{-2^i}\right)^t=\left[\sum_{i=0}^{D-1}z^{-i}\right]^t=\left[\frac{1-z^{-D}}{1-z^{-1}}\right]^t
\end{equation}
whereby the first equality holds for any $D$ that can be written
as an integer power of $2$, i.e., $D=2^p$. On the other hand, the
last equality holds for any integer value of $D$. Notice that
decimation factors of the form $2^p$ are quite common in practice.
Upon using~(\ref{formula_comb_1_ordine_2}) with $t=3$ and $D=2^3$,
(\ref{H16_2_z}) can be rewritten as follows:
\begin{equation}
\label{H16_2_z_recursive}
\begin{array}{lll}
H_{8,2}(z)&=&\left[\prod_{i=0}^{2}\left(1+z^{-2^i}\right)^3\right]\frac{1}{1+z^{-1}}\nonumber\\
&=&\left[\frac{1-z^{-8}}{1-z^{-1}}\right]^3\frac{1}{1+z^{-1}}
%
\end{array}
\end{equation}
%
%
%
%
The last relation in~(\ref{H16_2_z_recursive}) can be simplified
as follows:
\begin{equation}\label{H16_3_z_recursive}\small
\frac{(1-z^{-8})^3}{(1-z^{-1})^3(1+z^{-1})}=\frac{(1-z^{-8})^3}{1-2z^{-1}+2z^{-3}-z^{-4}}
\end{equation}
A recursive implementation of filter $H_{8,2}(z)$ in
(\ref{H16_3_z_recursive}) is shown in Fig.~\ref{n_rec_impl_gcf3}e.
It is obtained in the same way as for a classic cascade
integrator-comb (CIC) implementation \cite{Hoge}. In other words,
the numerator in~(\ref{H16_3_z_recursive}) corresponds to the comb
sections at the right of the decimator by\footnote{Notice that
$(1-z^{-8})^3$ becomes $(1-z^{-1})^3$ upon its shifting through
the decimator by $D=8$.} $D$, while the denominator is responsible
for the integrator sections at the left of the decimator by $D=8$.

The derivations yielding~(\ref{H16_3_z_recursive}) upon starting
from~(\ref{H16_2_z_recursive}) can also be accomplished by
following another reasoning\footnote{We discuss this other
approach for completeness, since it can be effective for deriving
an appropriate architecture for other decimation filter shown in
Table~\ref{Risultati_opt}.} based on the following relation:
\begin{equation}\label{cyclo_1_xn}
    1+z^{-n}=\frac{1-z^{-2n}}{1-z^{-n}}
\end{equation}
which is valid for any positive $n=2^{t-1}w$ with $w$ an odd
integer. By doing so,~(\ref{H16_2_z}) can be rewritten as follows:
\begin{eqnarray}\label{h82z_another}
H_{8,2}(z)&=&\left(1+z^{-1}\right)^2\left(1+z^{-2}\right)^3\left(1+z^{-4}\right)^3\\
&=&\left(\frac{1-z^{-2}}{1-z^{-1}}\right)^2\left(\frac{1-z^{-4}}{1-z^{-2}}\right)^3\left(\frac{1-z^{-8}}{1-z^{-4}}\right)^3\nonumber
\end{eqnarray}
Upon simplifying,~(\ref{h82z_another})
yields~(\ref{H16_3_z_recursive}).

An alternative non recursive architecture stems from a full
polyphase decomposition of the transfer function $H_{8,2}(z)$.
Upon solving polynomial multiplications in~(\ref{H16_2_z}),
$H_{8,2}(z)$ can be rewritten as follows:
\figura{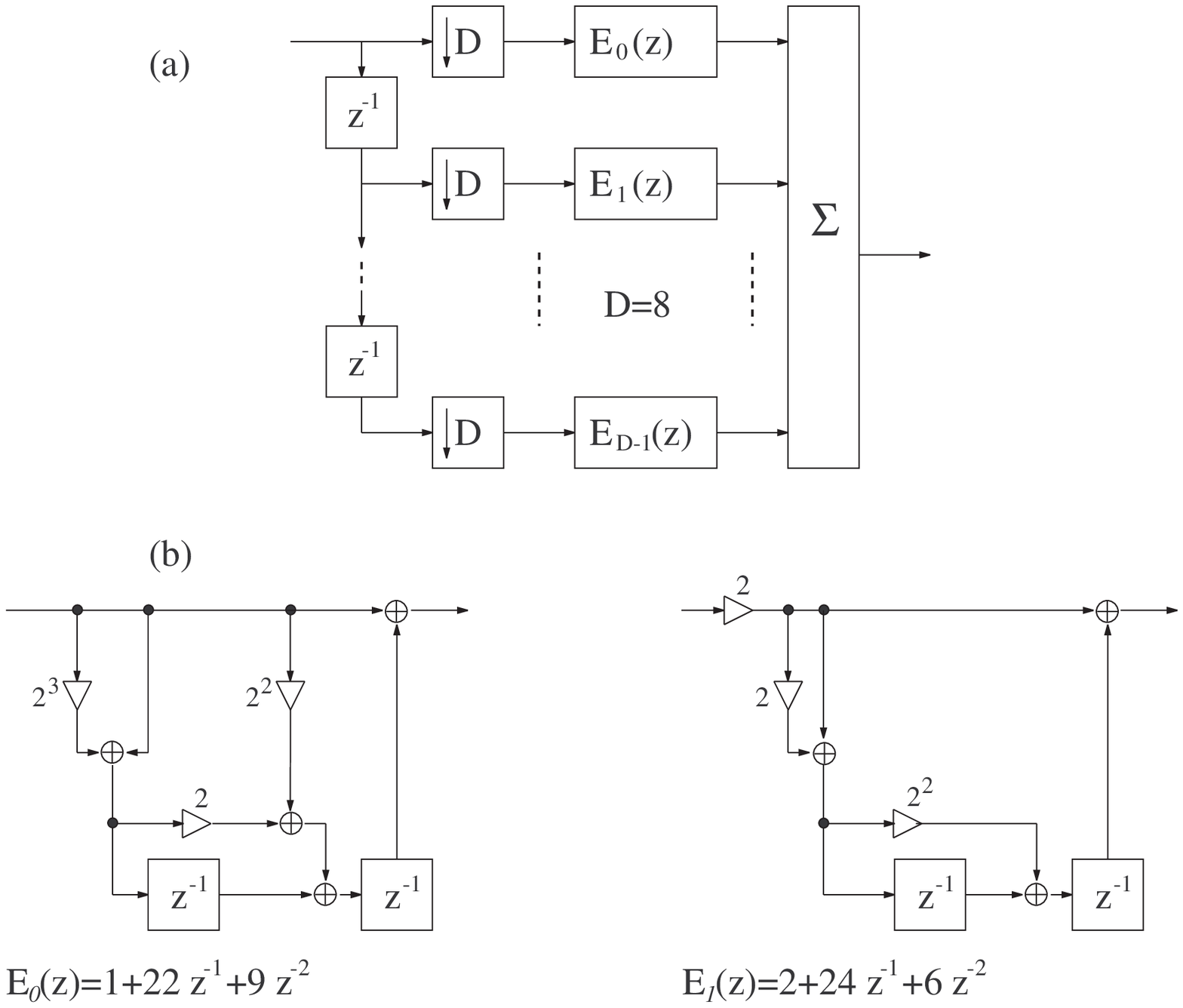}{Architecture of the polyphase
implementation of the decimation filter $H_{8,2}(z)$ (a), and
efficient design of the first two polyphase components $E_0(z)$
and $E_1(z)$ (b).}{polyphase_decomposition}
\begin{equation}\label{hd162_formaestesa}
H_{8,2}(z)=\sum_{i=0}^{20}h(i)z^{-i}
\end{equation}
By applying the polyphase decomposition \cite{antoniou},
$H_{8,2}(z)$ can be rewritten as
\begin{eqnarray}\label{hd162_formaestesa_polydec}
H(z)&=&\sum_{i=0}^{D-1}z^{-i}E_i(z^D)\\
E_i(z)&=&\sum_{t=0}^{\left\lfloor\frac{L}{D}\right\rfloor}h(D\cdot
t+i)z^{-t},~0\le i\le D-1\nonumber
\end{eqnarray}
whereby $L=21$ is the length of the impulse response. The
$z$-transfer function in~(\ref{hd162_formaestesa_polydec}) is
implemented with the architecture shown in
Fig.~\ref{polyphase_decomposition}a. The polyphase components
$E_i(z),~0\le i\le D-1=7,$ can be easily obtained by
employing~(\ref{hd162_formaestesa}). In particular, the first two
polyphase components take on the following expressions:
\begin{equation}\label{hd162_formaestesa_pc}
\begin{array}{lll}
E_0(z)&=&1+22z^{-1}+9z^{-2}\\
&=&(2^3+2^0)z^{-1}(2+z^{-1})+2^2z^{-1}+1\\
E_1(z)&=&2+24z^{-1}+6z^{-2}\\
&=&2\left[1+(2^1+1)z^{-1}(z^{-1}+2^2)\right]
\end{array}
\end{equation}
An efficient architecture for implementing each polyphase
component $E_i(z)$ stems from the decomposition of each integer as
the summation of power-of-two coefficients as shown
in~(\ref{hd162_formaestesa_pc}) for the first two polyphase
components $E_0(z)$ and $E_1(z)$. By doing so, and employing
coefficient sharing arguments, practical architectures featuring a
minimum number of shift registers easily follow as depicted in
Fig.~\ref{polyphase_decomposition}b. Similar considerations can be
employed for obtaining the architectures of the remaining
polyphase components $E_2(z),\ldots,E_{7}(z)$.
\section{Conclusions}
\label{conclusions}
This paper addressed the design of multiplier-less decimation
filters suitable for oversampled digital signals. The aim was
twofold. On one hand, it proposed an optimization framework for
the design of constituent decimation filters in a general
multistage decimation architecture using as basic building blocks
cyclotomic polynomials (CPs), since the first 104 CPs have simple
coefficients ($\{-1,0,+1\}$). On the other hand, the paper
provided a bunch of useful techniques, most of which stemming from
some key properties of CPs, for designing the optimized filters in
a variety of architectures. Both recursive and non-recursive
architectures have been discussed by focusing on a specific
decimation filter obtained as a result of the optimization
algorithm. Design guidelines were provided with the aim to
simplify the design of the constituent decimation filters in the
multistage chain.
\begin{table*}
\caption{The first sixty cyclotomic polynomials.}
\begin{center}
\begin{tabular}{l|l||l|l||l|l} \hline

$q$&$C_q(z^{-1})$ & $q$&$C_q(z^{-1})$ & $q$&$C_q(z^{-1})$
\\\hline \hline

1 & $1-z^{-1}$ & 11 &
$\sum_{i=0}^{10}z^{-i}=\frac{1-z^{-11}}{1-z^{-1}}$ & 21 & $1-z^{-1}+z^{-3}-z^{-4}+z^{-6}$\\

 &  &  &
&  & $-z^{-8}+z^{-9}-z^{-11}+z^{-12}$\\

 &  &  &
&  & $=\frac{1+z^{-7}+z^{-14}}{1+z^{-1}+z^{-2}}$\\

2 & $1+z^{-1}$ & 12 & $1-z^{-2}+z^{-4}=\frac{1+z^{-6}}{1+z^{-2}}$ & 22 & $\sum_{i=0}^{10}(-1)^i z^{-i}=\frac{1+z^{-11}}{1+z^{-1}}$\\
3 & $1+z^{-1}+z^{-2}=\frac{1-z^{-3}}{1-z^{-1}}$ &  13 &
$\sum_{i=0}^{12}z^{-i}=\frac{1-z^{-13}}{1-z^{-1}}$ & 23 &$\frac{1-z^{-23}}{1-z^{-1}}$ \\
4 & $1+z^{-2}$ & 14 & $\sum_{i=0}^{6}(-1)^i z^{-i}=\frac{1+z^{-7}}{1+z^{-1}}$ & 24 &$1-z^{-4}+z^{-8}=\frac{1+z^{-12}}{1+z^{-4}}$\\
5 & $\frac{1-z^{-5}}{1-z^{-1}}$& 15
& $1-z^{-1}+z^{-3}-z^{-4}+z^{-5}$ &25 &$\sum_{i=0}^{4}z^{-5i}=\frac{1-z^{-25}}{1-z^{-5}}$\\
 & &
& $-z^{-7}+z^{-8}=\frac{1+z^{-5}+z^{-10}}{1+z^{-1}+z^{-2}}$ & &\\

6 & $1-z^{-1}+z^{-2}=\frac{1+z^{-3}}{1+z^{-1}}$& 16& $1+z^{-8}$ &26 &$\sum_{i=0}^{12}(-1)^i z^{-i}=\frac{1+z^{-13}}{1+z^{-1}}$\\
7 & $\frac{1-z^{-7}}{1-z^{-1}}$& 17 & $\sum_{i=0}^{16}z^{-i}=\frac{1-z^{-17}}{1-z^{-1}}$ & 27 &$1+z^{-9}+z^{-18}=\frac{1-z^{-27}}{1-z^{-9}}$\\
8 & $1+z^{-4}$& 18& $1-z^{-3}+z^{-6}=\frac{1+z^{-9}}{1+z^{-3}}$& 28&$\sum_{i=0}^{6}(-1)^i z^{-2i}=\frac{1+z^{-14}}{1+z^{-2}}$\\
9 & $1+z^{-3}+z^{-6}=\frac{1-z^{-9}}{1-z^{-3}}$& 19&$\sum_{i=0}^{18}z^{-i}=\frac{1-z^{-19}}{1-z^{-1}}$&29&$\frac{1-z^{-29}}{1-z^{-1}}$\\
10 & $\sum_{i=0}^{4}(-1)^i z^{-i}=\frac{1+z^{-5}}{1+z^{-1}}$&20&$1-z^{-2}+z^{-4}-z^{-6}+z^{-8}$&30&$1+z^{-1}-z^{-3}-z^{-4}-z^{-5}$\\
 & &&$=\frac{1+z^{-10}}{1+z^{-2}}$&&$+z^{-7}+z^{-8}=\frac{1-z^{-5}+z^{-10}}{1-z^{-1}+z^{-2}}$\\\hline

31 & $\frac{1-z^{-31}}{1-z^{-1}}$& 41&
$\frac{1-z^{-41}}{1-z^{-1}}$ &51 & $1 - z^{-1} + z^{-3} - z^{-4} +
z^{-6} - z^{-7} $ \\
 & & &
 & & $ z^{-9} - z^{-10} + z^{-12} - z^{-13} +
            z^{-15}$ \\

 & & &
 & & $- z^{-16} + z^{-17} - z^{-19} + z^{-20} - z^{-22} $ \\

 & & &
 & & $ z^{-23} - z^{-25} +
            z^{-26} - z^{-28} +z^{-29} $ \\
 & & &
 & & $  - z^{-31} + z^{-32}=\frac{1-z^{-51}}{1-z^{-3}}\cdot \frac{1-z^{-1}}{1-z^{-17}}$ \\

32 &$1+z^{-16}$ & 42& $1 + z^{-1} - z^{-3} - z^{-4} + z^{-6} $  &52 &$\sum_{i=0}^{12}(-1)^i z^{-2i}=\frac{1+z^{-26}}{1+z^{-2}}$\\

 & & & $-z^{-8} - z^{-9} +
z^{-11} + z^{-12}$  & &\\

 & & & $=\frac{1-z^{-7}+z^{-14}}{1-z^{-1}+z^{-2}}$  & &\\

33 & $1 - z^{-1} + z^{-3} - z^{-4} + z^{-6} $   & 43& $\frac{1-z^{-43}}{1-z^{-1}}$ &53 &$\frac{1-z^{-53}}{1-z^{-1}}$\\

 & $-z^{-7} + z^{-9} -
z^{-10} + z^{-11} $& &  & & \\

&$- z^{-13} +
    z^{-14} - z^{-16} + z^{-17} $& &  & & \\

&$- z^{-19} + z^{-20}=\frac{1+z^{-11}+z^{-22}}{1+z^{-1}+z^{-2}}$& &  & & \\

34 &$\sum_{i=0}^{16}(-1)^i z^{-i}=\frac{1+z^{-17}}{1+z^{-1}}$ & 44& $\sum_{i=0}^{10}(-1)^i z^{-2i}=\frac{1+z^{-22}}{1+z^{-2}}$ &54 & $1-z^{-9}+z^{-18}$\\

35 & $1 - z^{-1} + z^{-5} - z^{-6} + z^{-7} $ & 45& $1 - z^{-3} +
z^{-9} - z^{-12} $  &55 &  $1 - z^{-1} + z^{-5} - z^{-6} +
z^{-10}- z^{-12} $ \\

& $ - z^{-8} + z^{-10} - z^{-11} + z^{-12}  $ & & $ + z^{-15} -
z^{-21} + z^{-24}$  && $z^{-15} - z^{-17} + z^{-20} -
            z^{-23} + z^{-25} $\\

& $- z^{-13} + z^{-14} - z^{-16} + z^{-17}  $ & & $=\frac{1+z^{-15}+z^{-30}}{1+z^{-3}+z^{-6}}$ && $ - z^{-28} + z^{-30} - z^{-34} + z^{-35} $\\

& $- z^{-18} + z^{-19} - z^{-23} + z^{-24}
       $ & &  && $- z^{-39} + z^{-40}=\frac{1-z^{-55}}{1-z^{-5}}\cdot \frac{1-z^{-1}}{1-z^{-11}}$\\

& $=\frac{1-z^{-1}-z^{-35}+z^{-36}}{1-z^{-5}-z^{-7}+z^{-12}}$ & &  &&\\

36 & $1 - z^{-6} + z^{-12}$& 46& $\sum_{i=0}^{22}(-1)^i z^{-i}=\frac{1+z^{-23}}{1+z^{-1}}$ &56 & $\sum_{i=0}^{6}(-1)^i z^{-4i}=\frac{1+z^{-28}}{1+z^{-4}}$\\

37 & $\frac{1-z^{-37}}{1-z^{-1}}$& 47&
$\frac{1-z^{-47}}{1-z^{-1}}$ &57 & $1 - z^{-1} + z^{-3} - z^{-4} +
z^{-6} - z^{-7}  $\\

& & &  & & $+ z^{-9}- z^{-10} + z^{-12} - z^{-13} $\\

& & &  & & $ +
            z^{-15} - z^{-16}+z^{-18} - z^{-20} + z^{-21}  $\\

& & &  & & $ - z^{-23} + z^{-24} -
z^{-26}+z^{-27}- z^{-29}  $\\

& & &  & & $ + z^{-30} - z^{-32} + z^{-33} - z^{-35} + z^{-36}$\\

& & &  & & $ =\frac{1-z^{-57}}{1-z^{-3}}\cdot \frac{1-z^{-1}}{1-z^{-19}}$\\

38 &$\sum_{i=0}^{18}(-1)^i z^{-i}=\frac{1+z^{-19}}{1+z^{-1}}$ & 48& $1-z^{-8}+z^{-16}$ &58 &$\sum_{i=0}^{28}(-1)^i z^{-i}=\frac{1+z^{-29}}{1+z^{-1}}$ \\

39 & $1 - z^{-1} + z^{-3} -z^{-4} + z^{-6} $  & 49&$\sum_{i=0}^{6} z^{-7i}=\frac{1-z^{-49}}{1-z^{-7}}$   &59 &$\frac{1-z^{-59}}{1-z^{-1}}$ \\

 & $- z^{-7} + z^{-9}-
z^{-10}+ z^{-12} $  & & & & \\

 & $- z^{-14} +
            z^{-15}- z^{-17} + z^{-18} $  & & & & \\

 & $ - z^{-20} + z^{-21} - z^{-23}+ z^{-24}$  & & & & \\

 & $=\frac{1- z^{-39}}{1- z^{-3}}\cdot\frac{1- z^{-1}}{1- z^{-13}}$  & & & & \\

40 &$\sum_{i=0}^{4}(-1)^i z^{-4i}=\frac{1+z^{-20}}{1+z^{-4}}$ &
50& $\sum_{i=0}^{4}(-1)^i
z^{-5i}=\frac{1+z^{-25}}{1+z^{-5}}$ &60 & $1 + z^{-2} - z^{-6} - z^{-8} - z^{-10} $\\

 & & &  & & $+ z^{-14} +
z^{-16}=\frac{1-z^{-10}+z^{-20}}{1-z^{-2}+z^{-4}}$\\

\hline

\hline

\hline\hline

\end{tabular}
\label{some_cyclot_polyn}
\end{center}
\end{table*}
\end{document}